\documentclass[12pt]{article}
\usepackage[usenames,dvipsnames]{color}
\usepackage{amsmath,amsthm,amsfonts,amssymb,multicol,amscd,amsbsy}
\usepackage{graphicx}
\usepackage{booktabs}
\usepackage{array}
\usepackage{enumerate}
\usepackage{url}
\usepackage[nodayofweek]{datetime}
\usepackage[english]{babel}
\usepackage{mathtools}
\usepackage[toc,page]{appendix}
\usepackage{verbatim} 
\usepackage{amsthm} 
\usepackage{enumitem}
\usepackage{enumerate}
\usepackage{bbm} 
\usepackage[normalem]{ulem} 
\usepackage{bm}
\usepackage{hyperref}
\usepackage{latexsym}
\usepackage[font=footnotesize,labelfont=bf]{caption}

\usepackage{tikz}

\usetikzlibrary{decorations.pathreplacing,decorations.markings}
\usepackage{subfig}
\usepackage{xcolor}
\usepackage{xspace}

\tikzset{
  mid arrow/.style={postaction={decorate,decoration={
        markings,
        mark=at position .5 with {\arrow[#1]{stealth}}
      }}},
}

\newcommand{\hee}{\tikz{\draw[thick] (0,0)--(0.2,0)--(0.4,0);\draw[fill=black] (0,0) circle (1pt);\draw[fill=black] (0.2,0) circle (1pt);\draw[fill=black] (0.4,0) circle (1pt);}}
\newcommand{\hve}{\tikz{\draw[thick] (0,0)--(0.2,0)--(0.2,-0.2);\draw[fill=black] (0,0) circle (1pt);\draw[fill=black] (0.2,0) circle (1pt);\draw[fill=black] (0.2,-0.2) circle (1pt);}}

\usepackage{authblk}

\newcommand{\be}{\begin}
\newcommand{\e}{\end}
\newcommand{\beq}{\begin{equation}}
\newcommand{\eeq}{\end{equation}}

\renewcommand{\l}{\left}
\renewcommand{\r}{\right}

\newcommand{\set}[1]{\mathbb{#1}}
\newcommand{\curly}[1]{\mathcal{#1}}

\newcommand{\setof}[2]{\left\{ #1\; : \;#2 \right\}}

\newcommand{\R}{\set{R}}
\newcommand{\C}{\set{C}}
\newcommand{\Z}{\set{Z}}

\newcommand{\eps}{\epsilon}

\newcommand{\Lam}{\Lambda}
\newcommand{\gam}{\gamma}

\theoremstyle{definition}

\numberwithin{equation}{section}

\theoremstyle{remark}

\def\dotuline{\bgroup
  \ifdim\ULdepth=\maxdimen  
   \settodepth\ULdepth{(j}\advance\ULdepth.4pt\fi
  \markoverwith{\begingroup
  \advance\ULdepth0.08ex
  \lower\ULdepth\hbox{\kern.15em .\kern.1em}%
  \endgroup}\ULon}

\def\dashuline{\bgroup
  \ifdim\ULdepth=\maxdimen  
   \settodepth\ULdepth{(j}\advance\ULdepth.4pt\fi
  \markoverwith{\kern.15em
  \vtop{\kern\ULdepth \hrule width .3em}%
  \kern.15em}\ULon}
\allowdisplaybreaks

\begin{document}
\title{Finite-size criteria for spectral gaps in $D$-dimensional quantum spin systems}
\author{Marius Lemm}
\affil{\textit{Department of Mathematics, Harvard University, Cambridge, MA 02138}}
\date{February 19, 2019}

\maketitle
\begin{abstract}
We generalize the existing finite-size criteria for spectral gaps of frustration-free spin systems to $D>2$ dimensions. We obtain a local gap threshold of $\frac{3}{n}$, independent of $D$, for nearest-neighbor interactions. The $\frac{1}{n}$ scaling persists for arbitrary finite-range interactions in $\Z^3$. The key observation is that there is more flexibility in Knabe's combinatorial approach if one employs the operator Cauchy-Schwarz inequality.
\end{abstract}

\section{Introduction}
A fundamental question concerning any quantum spin system is whether the Hamiltonian operator defining it is \emph{gapped} or \emph{gapless}. (We say that a Hamiltonian is gapped, if the spectral gap between its lowest and second-lowest eigenvalues is bounded away from zero in the thermodynamic limit of large system size. Otherwise, it is gapless.)
  
The existence of a spectral gap has far-reaching consequences for the low-energy physics of the system. For example, it is known that ground states of gapped Hamiltonians are well-controlled; they exhibit exponential clustering \cite{NS}, and they satisfy various notions of finite complexity \cite{A1,A2,A3}. (The latter fact is only proved rigorously in one dimension. Establishing it in higher dimensions is a major open problem.) Moreover, the closing of the spectral gap, as a system parameter is varied, indicates the occurrence of a quantum phase transition. 

Despite the central importance of spectral gaps, the mathematical methods for deriving them are somewhat limited.  The two main techniques for deriving a spectral gap are (a) the martingale method \cite{Nachtergaele} and (b) finite-size criteria. The main shortcoming of both of these methods is that they are limited to the special class of frustration-free quantum spin systems (see Assumption \ref{ass:FF1D} below). We also mention that the undecidable halting problem can be embedded as the question whether a certain translation-invariant Hamiltonian is gapped \cite{BCLP18,CPW}.

In this paper, we focus on combinatorial finite-size criteria as first established in a paper by Knabe \cite{K}, which was inspired by the proof that the AKLT chain (named after Affleck, Kennedy, Lieb and Tasaki) is gapped \cite{AKLT}. (We mention that there exists an alternative finite-size criterion due to \cite{FNW} which is used, e.g., in the recent work \cite{Aetal}.) \\

Our main contribution here is to show that the combinatorial finite-size criteria {\`a} la Knabe can be extended to $D$-dimensional frustration-free quantum spin systems, where $D$, the dimension of the underlying lattice $\Z^D$, can be arbitrary. Previous results \cite{GM,K,LM} are restricted to one and two dimensions for a technical reason which we resolve here by controlling certain operator-valued correction terms using the operator Cauchy-Schwarz inequality.
 
The main result is a finite-size criterion for nearest-neighbor interactions in $D$ dimensions, with an explicit gap threshold $\leq \frac{3}{n}$, where $n$ is the linear size $n$ of the subsystem (Theorem \ref{thm:3D}). This slightly improves upon an asymptotic gap threshold $o\l(\frac{(\log n)^{2+\eps}}{n}\r)$ established in \cite{KLucia} via the martingale method. We also explain how to extend the argument to arbitrary finite-range interactions in $3$ dimensions using the coarse-graining technique developed in \cite{LM}; see Theorem \ref{thm:FR}. Throughout, we work with periodic boundary conditions for simplicity. 

We view our main contribution to be a ``proof of principle'' that the finite-size techniques generally apply in arbitrary dimensions and yield a non-trivial scaling of the gap threshold. We leave it as an open problem to improve the gap threshold scaling to $o(n^{-1})$, like the $n^{-2}$ or $n^{-3/2}$ scaling established in \cite{GM,LM}, respectively, depending on the boundary conditions.  Such an improvement is likely possible using the weighting method as in \cite{GM,LM}, but we leave this to future work. We emphasize that such an improvement would not just be an academic fact: As observed in \cite{LM}, a gap threshold scaling of $o(n^{-1})$ for open boundary conditions would exclude the presence of chiral massless edge modes in frustration-free systems in any dimension $D>2$ (whereas \cite{LM} considered the case $D=2$).

Before we discuss these new results, we review the existing combinatorial finite-size criteria {\`a} la Knabe for $D\in\{1,2\}$.

\section{Review of existing finite-size criteria {\`a} la Knabe}
For simplicity, we focus on the finite-size criteria in one dimension from \cite{GM,LM}, and we comment on their two-dimensional analogs along the way. The Hilbert space of a chain of $m$ quantum spins is
$$
\curly{H}_m:=\bigotimes_{j=1}^m \C^d.
$$
For convenience, we restrict to translation-invariant systems so that the interaction is given by a fixed projection matrix $P:\C^d\otimes \C^d\to \C^d\otimes \C^d$, which we embed into the set of operators on $\curly{H}_m$ in the usual way. Namely, given $1\leq j\leq m$, we set
$$
h_{j,j+1}:=P\otimes \mathrm{Id}_{\{1,\ldots,m\}\setminus\{j,j+1\}}.
$$
(We implicitly compute modulo $m$, i.e., $h_{m,m+1}\equiv h_{m,1}$.) This defines a Hamiltonian with open boundary conditions
$$
H_m:=\sum_{j=1}^{m-1} h_{j,j+1},
$$
and one with periodic boundary conditions
$$
H_m^{per}:=H_m+h_{m,1}.
$$
Notice that since $P\geq 0 $ is a projection, we also have $H_m,H_m^{per}\geq 0$. From the perspective of energy minimization, both of these operators express a series of (non-commuting) constraints defined by $\ker h_{j,j+1}$. The frustration-free class is the special class of quantum spin systems for which all constraints can be simultaneously satisfied. This is expressed by the following assumption.

\be{ass}[Frustration-free]\label{ass:FF1D}
We have
$$
\ker H_m^{per}\neq \{0\}.
$$
\e{ass}

Notice that this assumption automatically ensures that $\ker H_m\neq \{0\}$ as well. 

\be{rmk}
Deriving spectral gaps outside of the frustration-free class constitutes a major open problem, and is related to a famous conjecture of Haldane \cite{Haldane1,Haldane2} that the integer-spin antiferromagnetic Heisenberg chain is gapped. We will not have anything further to say on frustrated spin systems in the present work. 
\e{rmk}

\be{defn}
The spectral gap $\gam_m$ of $H_m$ is the smallest non-zero eigenvalue of $H_m$. We define $\gam_m^{per}$ analogously.
\e{defn}

 We emphasize that, by this convention, the spectral gap is always positive, even in the case of a degenerate ground state. With this definition at hand, we can make precise the notion of ``gapped Hamiltonian''. If there exists a constant $c>0$ such that $\gam_m\geq c>0$ holds for all $m\geq 2$, then we say that $H_m$ is gapped. Otherwise, it is gapless. We make an analogous definition for $H_m^{per}$ and $\gam_m^{per}$. 
 
 We are now ready to state the recent finite-size criteria for one-dimensional quantum spin systems from \cite{GM,LM}. Both of these references also contain finite-size criteria in two dimensions to different degrees of generality.

\be{thm}[\cite{GM}]\label{thm:GM}
Let $n>2$ and $m>2n$. Then
\beq\label{eq:GM}
\gam_m^{per}\geq \frac{5}{6}\frac{n^2+n}{n^2-4} \l(\gam_n-\frac{6}{n(n+1)}\r).
\eeq
\e{thm}

Notice that \eqref{eq:GM} involves the two different gaps $\gam_m^{per}$ and $\gam_n$. By construction, it functions as a finite-size criterion for showing that the periodic Hamiltonian $H_m^{per}$ is gapped. Indeed, suppose there exists a finite $n$ (say, $n=5$) such that the spectral gap $\gam_n$ exceeds the ``gap threshold'' $\frac{6}{n(n+1)}$. If this is the case, then \eqref{eq:GM} implies that $\gam_m^{per}\geq c>0$ for all $m\geq 2$ and hence $H_m^{per}$ is gapped. 

By contrast, the other recent one-dimensional finite-size criterion involves only the spectral gaps $\gam_n$, simultaneously at various comparable system sizes.

\be{thm}[\cite{LM}]\label{thm:LM}
Let $n>3$ and $m>2n$. Then
\beq\label{eq:LM}
\gam_m\geq \frac{1}{2^9\sqrt{6} n} \l(\min_{n/2\leq \ell\leq n}\gam_\ell-\frac{4\sqrt{6}}{n^{3/2}}\r).
\eeq
\e{thm}

This theorem functions as a finite-size criterion to show that the Hamiltonian $H_m$ is gapped, i.e., for spin chains with an open boundary. Indeed, in a completely analogous way as described before, if one can show that $\min_{n/2\leq \ell\leq n}\gam_\ell$ exceeds the gap threshold $\frac{4\sqrt{6}}{n^{3/2}}$ at some fixed, finite $n$, then the whole Hamiltonian $H_m$ is gapped. 
Both Theorems \ref{thm:GM} and \ref{thm:LM} also imply statements about systems with a sufficiently slow gap closing rate. This ``converse version'' was used to classify spin $1/2$ chains \cite{BGos}. Here we focus on consequences of Theorem \ref{thm:LM}. Since it involves the spectral gap of the same Hamiltonian on both sides of the inequality, we can use it to infer the following corollary.

\be{cor}[of Theorem \ref{thm:LM}]
If $H_n$ is gapless, then $\min_{n/2\leq \ell\leq n}\gam_\ell=O(n^{-3/2})$.
\e{cor}

Essentially, this says that the ``gap cannot close too slowly''  in frustration-free spin chains with a boundary. This result and its two-dimensional analog in \cite{LM}  improve a recent bound from \cite{KLucia} in one and two dimensions. The main application is that it rigorously confirms the physics folklore that chiral edge modes cannot occur in 2D frustration-free systems. The reason is that chiral edge modes would have spectral gaps scaling like $n^{-1}$; see \cite{LM} for the details.

We collect some further remarks concerning Theorems \ref{thm:GM} and \ref{thm:LM}.

\be{rmk}
\be{enumerate}[label=(\roman*)]

\item To prove Theorems \ref{thm:GM} and \ref{thm:LM}, the key idea, which is originally due to Knabe \cite{K}, is to relate the squared Hamiltonian $H_m^2$ to the sum of squares of local Hamiltonians using combinatorial arguments.

\item Notice that the gap threshold scales like $n^{-3/2}$ in Theorem \ref{thm:LM} and like $n^{-2}$ in Theorem \ref{thm:GM}. The latter scaling is optimal, as can be seen from the ferromagnetic Heisenberg chain \cite{GM}. It is an open problem if the $n^{-3/2}$ scaling in Theorem \ref{thm:LM} can be improved to $n^{-2}$. This touches upon the delicate (and not fully understood) relation between boundary conditions and spectral gaps. In this context, we mention that $n^{-3/2}$ is precisely Kardar-Parisi-Zhang gap scaling behavior and can be observed in one-dimensional spin chains in which specific non-self-adjoint boundary terms effectively produce an interface \cite{deGier}.

\item As mentioned before, Theorem \ref{thm:GM} and \ref{thm:LM} also have two-dimensional analogs in \cite{GM,LM}. A crucial feature of these is that the two-dimensional subsystems (called ``patches'') have to be carefully designed so that a certain combinatorial qualification is satisfied (all the different kinds of ways to pair to adjacent edges need to appear the same number of times). It is precisely this restriction which is difficult to ensure in $D\geq 3$ and which we overcome here by using the operator Cauchy-Schwarz inequality.


\e{enumerate}
\e{rmk}

\section{Main results}
\subsection{Setup for $D$-dimensional spin systems}
We move from the setup for one-dimensional quantum spin chains to the $D$-dimensional case. To avoid confusion between these cases, we now use $N$ for the linear size of the whole system.

We let
$$
\Lam_N:=((-N,N]\cap \Z)^D
$$
be a periodic box in $\Z^D$. At each site of $\Lam_N$, we place a quantum spin of local dimension $d$ (so $d=2S+1$ where $S$ is the spin number). The Hilbert space is
$$
\curly{H}_{\Lam_N}:=\bigotimes_{j\in \Lam}\C^d.
$$

We will consider two different types of translation-invariant Hamiltonian. They are defined either by (a) nearest-neighbor interactions and arbitrary $D$, or (b) arbitrary finite-range interactions and $D=3$. We work with periodic boundary conditions for simplicity; the techniques from \cite{LM} can be used to extend the results to other (in particular, free) boundary conditions.

Given a pair of edges $j,k\in\Lam_N$, we write $j\sim k$ if they are connected by an edge in $\Z^3$, where we use periodic boundary conditions on $\Lam_N$.
Similarly to the one-dimensional case, we define the nearest-neighbor interactions
$$
h_{j,k}:=h\otimes \mathrm{Id}_{\Lam_N\setminus\{j,k\}}
$$
and the Hamiltonian
$$
H_N:=\sum_{\substack{j,k\in \Lam_N\\ j\sim k}} h_{j,k}.
$$
We write $\gam_N$ for its spectral gap.

\subsection{Main result for nearest-neighbor interactions}
The finite-size criterion for the nearest-neighbor interactions is easier to state and more practical. The subsystems we use are the boxes
$$
\curly{B}_n:=([0,n]\cap\Z)^D,
$$
considered as subgraphs of $\Z^D$. (This means that $\curly{B}_n$ is a box with open boundary conditions, whereas the whole system $\Lam_N$ has periodic boundary conditions.)

Given two sites $j,k\in \curly{B}_n$, we write $j\sim_{\curly{B}_n} k$ if $j$ and $k$ are connected by an edge in $\curly{B}_n$. The corresponding Hamiltonian is
$$
H_{\curly{B}_n}:=\sum_{\substack{j,k\in\curly{B}_n\\ j\sim_{\curly{B}_n} k}}h_{j,k}
$$
and we write $\gam_{\curly{B}_n}$ for its spectral gap. We emphasize that $H_{\curly{B}_n}$ has open boundary conditions, in contrast to $H_N$.

The basic idea we follow (originally due to Knabe \cite{K} in D=1 and 2) is to build up the full Hamiltonian $H_N$ from various translates of the subsystem Hamiltonian $H_{\curly{B}_n}$. 

Our main result reads as follows.

\be{thm}[Main result]\label{thm:3D}
Let $D\geq 3, n\geq 3$ and $N\geq 2n+1$. Assume that $H_N$ is frustration-free, i.e., $\ker H_N\neq \{0\}$. Then, we have the gap bound
\beq\label{eq:main}
\gam_N\geq \gam_{\curly{B}_n}-\frac{1}{n}-\frac{2}{n^2}.
\eeq
\e{thm}

This theorem functions as a finite-size criterion to show that $H_N$ is gapped. The criterion applies if, for any fixed $n$, the spectral gap $\gam_{\curly{B}_n}$ exceeds the gap threshold $\frac{1}{n}+\frac{2}{n^2}\leq \frac{3}{n}$. We find it surprising that the gap threshold is independent of $D$.

The criterion can potentially be applied to systems with small $n$ by exactly diagonalizing $H_{\curly{B}_n}$ and its converse also contains asymptotic information as presented in the following corollary.

\be{cor}
If $H_N$ is gapless, then $\gam_{\curly{B}_n}=O(n^{-1})$ as $n\to\infty$.
\e{cor}

It is likely that the methods of \cite{GM,LM} allow to improve the scaling of the gap threshold to a larger power in $n$; we have not attempted this here. 

We mention that the main technical novelty in the proof of Theorem \ref{thm:3D} is the observation that the problematic terms (called $Q$ below) may not satisfy the counting property that is usually required by Knabe's combinatorial method exactly, but the extent to which they fail to do so can be controlled. A crucial role in our proof is played by the operator Cauchy-Schwarz inequality in the following form for two projections:
$$
-h_{j,k}h_{j',k'}-h_{j',k'}h_{j,k}\leq (-h_{j,k})^2+h_{j',k'}^2=h_{j,k}+h_{j',k'}.
$$

\subsection{Setup for finite-range interactions in $D=3$}
This section is modeled after Sections 3.1 and 3.3 in \cite{LM} where the two-dimensional case was considered. 

The main message is that the method used to prove Theorem \ref{thm:3D} generalizes to arbitrary finite-range interactions by means of the one-step coarse-graining procedure introduced in \cite{LM}, but with constants that have to be controlled on a case-by-case bases. We focus on the case $D=3$ for simplicity.

We begin with the general setup for translation-invariant, finite-range interactions. We define a ``unit cell of interactions'' and translate it across $\Lam_N$. We fix a finite family $\curly{S}$ of subsets $S\subset\Lam_N$ containing the origin. For each $S\in \curly{S}$, we fix a projection $P^S:(\C^d)^{\otimes |S|}\to (\C^d)^{\otimes |S|}$, where $|S|$ is the cardinality of $S$. The set of $\{P^S\}_{S\in\curly{S}}$ now defines the unit cell of interactions.

Given a point $x\in \Lam_N$, we define the set
$$
x+S:=\setof{y\in\Lam_N}{y-x\in S}
$$
with periodic boundary conditions on the difference $y-x$. We define a projection operator $P^S_{x+S}:\curly{H}_{\Lam_N}\to \curly{H}_{\Lam_N}$ by 
$$
P^S_{x+S}:= P^S\otimes I_{\Lam_N\setminus (x+S)}.
$$
That is, $P^S_{x+S}$ acts non-trivially only on the subspace $\bigotimes_{y\in x+S}\C^d$ of the whole Hilbert space $\curly{H}_{\Lam_N}=\bigotimes_{y\in\Lam_N}\C^d$. 

\be{defn}
Let $e_1,e_2,e_3$ be the canonical basis of $\R^3$. We write $d_1(\cdot)$ for the $\ell^1$ distance on $\Z^3$, meaning,
$$
d_1\l(\sum_{k=1}^3a_ke_k,\sum_{k=1}^3b_ke_k\r):=\sum_{k=1}^3|a_k-b_k|.
$$
We write $\mathrm{diam}_1(A)$ for the diameter of a set $A\subset \Z^3$, taken with respect to $d_1$.
\e{defn}

\be{ass}[Finite interaction range]\label{ass:FR}
 There exists $R>0$ such that $\mathrm{diam}(S)<R$ for all $S\in\curly{S}$.
\e{ass}

We define the finite-range Hamiltonian as 
$$
H_N^{FR}:=\sum_{x\in \Lam_N}\sum_{S\in \curly{S}} P^S_{x+S}
$$
and we write $\gam_N^{FR}$ for its spectral gap.

\be{rmk}[Scope]
We emphasize that the Hamiltonian $H_N^{FR}$ in the finite-range case is effectively defined on any three-dimensional lattice,  not just $\Z^3$ (e.g., the bcc or fcc lattices), even as stated. This is because we only use $\Z^3$ to \emph{label the sites} of our lattice, while the edges of $\Z^3$ do not enter at all. Instead, the information about the interactions is contained in the interaction shapes $S\in\curly{S}$, and by defining appropriate interactions shapes, other three-dimensional lattices can be accommodated. The formalism also includes lattices with multiple points per unit cell, as one can form larger unit cells by suitably enlarging the dimension $d$ of the Hilbert space $\C^d$ at each site in $\Z^3$. \e{rmk}

For Theorem \ref{thm:FR}, we use the following subsystem Hamiltonians. Without loss of generality, we assume that the interaction range $R>0$ is an odd integer. Given $y\in \Z^3$, we write $C(y)$ for the cube of sidelength $R$, centered at $y$, i.e.,
$$
C(y):=\setof{y+\sum_{k=1}^3 a_ke_k}{|a_k|\leq (R-1)/2,\,\,\textnormal{for } k=1,2,3}.
$$
We will use the following analog of the boxes $\curly{B}_n$ from the nearest-neighbor case:
$$
\curly{C}_n:=\bigcup_{0\leq b_1,b_2,b_3\leq n} C\l(\sum_{k=1}^3 R b_k e_k\r)
$$
with the associated frustration-free subsystem Hamiltonian
$$
H_{\curly{C}_n}:=\sum_{x\in \curly{C}_n} \sum_{\substack{S\in\curly{S}\\ x+S\subset \curly{C}_n}}
P^S_{x+S}
$$
The constraint $x+S\subset \curly{C}_n$ implements open boundary conditions for $H_{\curly{C}_n}$. We write $\gam_{\curly{C}_n}$ for the spectral gap of $H_{\curly{C}_n}$.

\subsection{Main result for general finite-range interactions}
We state an analog of Theorem \ref{thm:3D}. It applies for general finite-range interactions, with a less explicit constant (but which only depends on the details of the model under consideration). 

\be{thm}\label{thm:FR}
 Assume that $H_{N}^{FR}$ is frustration-free for all $N\geq 1$. There exist constants 
$$
c_1=c_1(\{P^S\}_{S\in\curly{S}}),\qquad c_2=c_2(\{P^S\}_{S\in\curly{S}}),
$$
such that the following gap bound holds for any $n\geq 2$ and $N\geq 2n+1$:
\beq\label{eq:thm3D}
\gam_{N}^{FR}\geq c_1\l(\gam_{\curly{C}_n}-\frac{c_2}{n}\r).
\eeq
\e{thm}

The more important constant is $c_2$, since it functions as the gap threshold. It is in principle computable from the knowledge of the unit cell of interactions $\{P^S\}_{S\in\curly{S}}$ only.

\section{Proof of Theorem \ref{thm:3D}}
\subsection{Step 1: Squaring the Hamiltonian}
By frustration-freeness and the spectral theorem, the claim is equivalent to the operator inequality
\beq\label{eq:claim}
H_N^2\geq \l(\gam_{\curly{B}_n}-\frac{1}{n}-\frac{2}{n^2}\r)H_N.
\eeq
We introduce some notation. Let $j,j',k,k'\in \Lam_N$. If the undirected edges corresponding to the pairs $(j,k)$ and $(j',k')$ are distinct but overlap at a vertex, then we write $(j,k)\sim (j',k')$. If they are distinct and do not overlap, we write $(j,k)\not\sim (j',k')$.

We start by squaring the Hamiltonian and use $h_{j,k}^2=h_{j,k}$ to obtain
\beq\label{eq:1}
\begin{aligned}
H_N^2=&H_N+Q+R,\\
\textnormal{where }\, Q:=&\sum_{\substack{\textnormal{edge pairs}\\(j,k)\sim (j',k')}} \{h_{j,k},h_{j',k'}\},\qquad R:=\sum_{\substack{\textnormal{edge pairs}\\(j,k)\not\sim (j',k') }} \{h_{j,k},h_{j',k'}\}
\end{aligned}
\eeq
and we wrote $\{A,B\}=AB+BA$ for the anticommutator between two matrices $A,B$. Notice that each summand contributing to $R$ is a product of commuting positive-definite matrices $h_{j,k}h_{j',k'}$, and is therefore itself positive definite (so in particular $R\geq 0$).

\subsection{Step 2: The subsystem Hamiltonians}
For every site $l\in\Lam_N$, we define the shifted box
$$
B_l:=\curly{B}_n+l:=\setof{k\in \Lam_N}{k-l\in \curly{B}_n},
$$
where we use the periodic boundary conditions of $\Lam_N$ when taking the difference $k-l$. Then we can define the shifted subsystem Hamiltonian
$$
H_{B_l}:=\sum_{\substack{j,k\in B_l\\ j\sim_{B_l} k}}h_{j,k}.
$$ 
Notice that for every $l\in \Lam_N$, the operators $H_{B_l}$ and $H_{\curly{B}_n}$ are unitarily equivalent via translation, and so their spectral gaps agree, $\gam_{B_l}=\gam_{\curly{B}_n}$.

We consider the auxiliary operator
\beq\label{eq:2}
A:=\sum_{l\in \Lam_N}(H_{B_l})^2=\sum_{l\in \Lam_N}\l(H_{B_l}+Q_{B_l}+R_{B_l}\r),
\eeq
where $Q_{B_l},R_{B_l}$ are defined analogously to \eqref{eq:1}, except that the relevant edge pairs are now contained in $B_l$. 

The following key proposition compares $A$ to $H_N^2$.

\be{prop}
\label{prop:key}
We have the two operator inequalities
\begin{align}
\label{eq:prop1}
A\leq&\l(n(n+1)^{D-1}+ 2 (n+1)^{D-2}\r) H_N+n^2(n+1)^{D-2}(Q+R),\\
\label{eq:prop2}
A\geq&\,\, n(n+1)^{D-1}\gam_{\curly{B}_n}H_N.
\end{align}
\e{prop}

\be{proof}[Proof of Proposition \ref{prop:key}]
We first note that
\beq\label{eq:Hcount}
\sum_{l\in \Lam_N}H_{B_l}=n(n+1)^{D-1} H_N,
\eeq
since every edge appears in exactly $n(n+1)^{D-1} $ boxes $\{B_l\}_{l\in \Lam_N}$. 

There are two types of $Q$ terms: ones for which the two edges are aligned (we call these $\hee$ terms) and ones for which the two edges are pointing in different directions (we call these $\hve$ terms). We count that each $\hee$ term appears in $(n-1)(n+1)^{D-1}$ boxes and each $\hve$ term appears in $n^2(n+1)^{D-2}$ boxes. Finally, all terms in $R$, i.e., pairs of edges which do not touch, appears in at most $n^2(n+1)^{D-2}$ boxes. (A formal proof of this fact can be given by using the technique used to prove the two-dimensional autocorrelation lemmas in \cite{GM,LM} and setting all weights $=1$.) Since all the interactions $h_{j,k}\geq 0$, these combinatorial considerations imply the operator inequality
$$
\begin{aligned}
A\leq&\, n(n+1)^{D-1} H_N + (n-1)(n+1)^{D-1} Q_{\hee}+n^2 (n+1)^{D-2} (Q_{\hve}+R)\\
=& n(n+1)^{D-1} H_N -(n+1)^{D-2}Q_{\hee}+n^2 (n+1)^{D-2} (Q+R).
\end{aligned}
$$
Here we used that $Q=Q_{\hee}+Q_{\hve}$. The only problematic term remaining is the one $-(n+1)^{D-2}Q_{\hee}$. To control it, we use the operator Cauchy-Schwarz inequality, which says
\beq
-\{h_{j,k},h_{j',k'}\}=-h_{j,k}h_{j',k'}-h_{j',k'}h_{j,k}\leq (-h_{j,k})^2+h_{j',k'}^2=h_{j,k}+h_{j',k'}.
\eeq
Hence,
$$
\begin{aligned}
-Q_{\hee}\leq \sum_{\substack{\textnormal{\hee\,\, edge pairs}\\(j,k)\sim (j',k')}} \l(h_{j,k}+h_{j',k'}\r)\leq 2 H_N.
\end{aligned}
$$
We have thus shown that
$$
A\leq \l(n(n+1)^{D-1}+ 2(n+1)^{D-2}\r) H_N+n^2(n+1)^{D-2}(Q+R).
$$
This proves \eqref{eq:prop1}.\\

For \eqref{eq:prop2}, we use that, by the spectral theorem and frustration-freeness,
$$
(H_{B_l})^2\geq \gam_{B_l}H_{B_l}=\gam_{\curly{B}_n} H_{B_l},
$$
where the second step holds by translation invariance of the local gap. By \eqref{eq:Hcount}, we conclude that
$$
A\geq \gam_{\curly{B}_n}\sum_{l\in \Lam_N}H_{B_l}=n(n+1)^{D-1}\gam_{\curly{B}_n}H_N,
$$
which proves \eqref{eq:prop2}.
\e{proof}

\subsection{Step 3: Conclusion}
We now combine \eqref{eq:1} with Proposition \ref{prop:key} to obtain
$$
\begin{aligned}
H_N^2\geq& H_N+Q+R\\
\geq& H_N+\frac{A-\l(n(n+1)^{D-1}+ 2 (n+1)^{D-2}\r) H_N}{n^2(n+1)^{D-2}}\\
\geq& H_N+\frac{n(n+1)^{D-1}\gam_{\curly{B}_n}H_N-\l(n(n+1)^{D-1}+ 2 (n+1)^{D-2}\r) H_N}{n^2(n+1)^{D-2}}\\
\geq & \l(1+\gam_{\curly{B}_n}-\frac{n+1}{n}-\frac{2}{n^2}\r)H_N\\
=&\l(\gam_{\curly{B}_n}-\frac{1}{n}-\frac{2}{n^2}\r)H_N
\end{aligned}
$$
This proves \eqref{eq:claim} and hence Theorem \ref{thm:3D}.
\qed

\section{Proof sketch for Theorem \ref{thm:FR}}
We only sketch the general line of argumentation, since it is essentially a combination of the proof of Theorem \ref{thm:3D} with the techniques developed in Section 6 of \cite{LM}, modulo modifications to pass from the two-dimensional situation to the three-dimensional situation. The procedure consists of two main steps.


\be{itemize}
\item \emph{Step 1: Coarse-graining procedure.} This follows section 6 in \cite{LM}. The frustration-free Hamiltonian $H_N^{FR}$, which by assumption has interaction range $R>0$, is replaced by a coarse-grained nearest-neighbor Hamiltonian $H_N^{cg}$. The ``metaspins'' on which $H_N^{cg}$ acts are exactly the boxes $C(y)$ and it only consists of three kinds of nearest-neighbor interactions. These arise from interactions between the $C(y)$-boxes that share either of the following: a corner; an edge; or a face.

The spectral gaps of the original Hamiltonian and of the coarse-grained Hamiltonian are related by constants that are uniform in the system size (these constants contribute to the constants $c_1,c_2$ in Theorem \ref{thm:FR}), thanks to the frustration-free assumption and the simple fact that a single box touches at most $12$ other boxes along an edge. (We remark that the analog of this step in higher dimensions would lead to a $D$-dependent constant.) From this point on, we work only with the coarse-grained Hamiltonian $H_N^{cg}$.
\item \emph{Step 2: Knabe-type argument for $H_N^{cg}$.} This follows the proof of Theorem \ref{thm:3D}, but now for the coarse-grained nearest-neighbor Hamiltonian $H_N^{cg}$. The first step is again to compute 
$$
(H_N^{cg})^2=H_N^{cg}+Q^{cg}+R^{cg}
$$
with $Q^{cg}$ and $R^{cg}$ defined analogously as in \eqref{eq:1}, i.e., $Q^{cg}$ contains exactly the non-commuting pairs of interactions. To obtain a lower bound on $Q^{cg}$, one considers the auxiliary operator
$$
A:=\sum_{x\in \Lam_N} H_{x+\curly{C}_n}^2,
$$
in analogy with \eqref{eq:2}. One then establishes an analog of Proposition \ref{prop:key} to relate $A$ to $(H_N^{cg})^2$. The main input are again combinatorial considerations and the operator Cauchy-Schwarz inequality to relate the count of the $3$ different kinds of $Q^{cg}$ terms (depending on whether two neighboring boxes touch at a corner, edge, or face). A crucial observation is that the difference in these $3$ counts is subleading by one order in $n$ (it scales as $n^2$ as opposed to the $n^3$ scaling of the main term), because the mismatch is only due to the boundary of $\curly{C}_n$.\e{itemize}

 In this way, one establishes a lower bound on the spectral gap of $H_N^{cg}$ in terms of $\gam_{\curly{C}_n}$. By step 1, one obtains a lower bound on the spectral gap $\gam_N^{FR}$ as well, and this proves Theorem \ref{thm:FR}.


\end{document}